\def\lsim{\raise0.3ex\hbox{$\;<$\kern-0.75em\raise-1.1ex
\hbox{$\sim\;$}}}
\def\gsim{\raise0.3ex\hbox{$\;>$\kern-0.75em\raise-1.1ex
\hbox{$\sim\;$}}}

\documentstyle[sprocl,psfig]{article}

\def\PRC{{\em Phys. Rev.} C}

\def\be{\begin{equation}}
\def\ee{\end{equation}}
\def\bea{\begin{eqnarray}}
\def\eea{\end{eqnarray}}
\begin{document}

%To Prof Nick Karayiannis -- do read this:-
%If needed the word of Chapter~1, you can type in at the 
%\title{}. The words will be in caps and lowercase. 
%For chapter title can be in all caps or in caps and lowercase.
%It is up to the author to type for the case sensitive but 
%all articles must be in the same style. 
%But mostly for Review Volume are without this Chapter~1.
%Thank you
%Jessie   13/4/2000

\title{
EXOTIC SOLUTIONS TO THE SOLAR NEUTRINO PROBLEM 
AND SOME IMPLICATIONS FOR LOW ENERGY SOLAR NEUTRINO
EXPERIMENTS
\footnote{Talk presented at International Workshop on Low 
Energy Solar Neutrinos (LowNu2), 4-5, December, 2000, Tokyo, Japan}
}

\author{HIROSHI NUNOKAWA}

\address{
Instituto de F\' {\i}sica Gleb Wataghin, 
     Universidade Estadual de Campinas, UNICAMP \\
     13083-970 Campinas SP, Brazil
\\E-mail: nunokawa@ifi.unicamp.br
}


\maketitle\abstracts{ 
In this talk, I review, from the phenomenological point of view, 
solutions to the solar neutrino problem, which are not 
provided by the conventional neutrino oscillation induced 
by mass and flavor mixing, and show that they can provide
a good fit to the observed data. 
I also consider some simple implications for low energy 
solar neutrino experiments. 
}

\section{Introduction}

It is considered that observed data coming from 
atmospheric neutrino experiments~\cite{atmospheric}
are compelling evidence of neutrino oscillation indicating 
the presence of neutrino mass and flavor mixing~\cite{MNS,pontecorvo}. 
This is now being confirmed by the ongoing K2K experiment~\cite{K2K}. 
Results of the solar neutrino 
experiments~\cite{solarnuexp}
are also supporting such mass induced neutrino oscillations hypothesis, 
either through the matter enhanced MSW mechanism~\cite{msw} or 
through the vacuum oscillation~\cite{pontecorvo}. 
I will call these explanations ``standard solutions''
to the solar neutrino problem (SNP)
as they are based only on neutrino mass and flavor mixing, 
the most natural extension of the standard model. 
Althogh not yet confirmed by other experiment, 
the LSND data~\cite{LSND} are also indicating neutrino 
mass and mixing. 

On the other hand, several alternative scenarios, 
which can explain these observations 
without invoking neutrino mass and/or flavor mixing, 
have been 
proposed~\cite{RSFP,GMP,fcsol2,roulet,wol,gasperini,atmexotic} 
and some of them are not yet excluded. 
I will call them ``non-standard'' or ``exotic'' solutions,
as they are theoretically less motivated compared to the 
standard solutions. 
In this talk, from the phenomenological point of view, 
I will review such non-standard solutions to the SNP 
which are provided by the neutrino conversion 
induced by resonant spin-flavor precession (RSFP)~\cite{RSFP}, 
non-standard neutrino interactions 
(NSNI)~\cite{GMP,fcsol2,roulet,wol},  
and violation of the equivalence principle (VEP)~\cite{gasperini}
and show that they can provide a good fit to the solar neutrino data. 
In the end, I will try to consider some possible implications
for low energy solar neutrino experiments. 
For non-standard explanations of the atmospheric 
neutrino observations, see Ref.~\cite{atmexotic} for a review. 

\newpage
\vspace{-0.15cm}
\section{Resonant Spin-Flavor Precession}
\vspace{-0.1cm}
Let me start with the solution to the SNP induced by 
neutrino magnetic moment, as I think it less exotic 
in the sense that this solution does require neutrino masses
in contrast to the other two solutions I will discuss 
in the next sections. 
If neutrinos have transition 
magnetic moment among different flavors, they 
can undergo spin-flavor precession (SFP)~\cite{schechtervalle}
in the presence of magnetic field. 
Moreover, in matter, such SFP
can be resonantly enhanced~\cite{RSFP} in the same fashion as in 
the case of the MSW effect~\cite{msw}. 
Such resonant conversion or RSFP can occur between 
$\nu_{eL} (\equiv \nu_e$) and $\nu^c_{\mu,\tau R} 
(\equiv {\bar{\nu}}_{\mu,\tau})$ for Majorana neutrinos and 
$\nu_{eL}$ and ${\nu}_{\mu,\tau R}$ for 
Dirac neutrinos
(${\nu}_{\mu,\tau R}$ is a electroweak singlet neutrino). 
In order to have RSFP conversion, neutrinos must
have different masses as in the case of the MSW effect, to satisfy 
the resonance condition~\cite{RSFP}. 

A nice feature of this mechanism is, as stressed in 
Ref.~\cite{LN_RSFP}, that it can give the appropriate 
energy dependent suppressions of neutrino fluxes 
needed to account well for the solar neutrino data, 
i.e., it can provide strong suppression for $^7$Be 
neutrinos, weak suppression for $pp$ neutrinos and 
moderate suppression for $^8$ B neutrinos. 
See Ref.~\cite{rsfp_recent} for recent analyses 
on this mechanism. 

Here I present some updated analysis of our previous 
work~\cite{GN_RSFP}, for the case of Majorana neutrinos
for $\nu_e \leftrightarrow \bar{\nu}_\mu$ 
(or $\bar{\nu}_\tau$) channel, 
taking one particular magnetic field profile 
used in our previous work~\cite{GN_RSFP}, which has 
the triangle shape concentrated in the solar convective zone
(profile 3 in Ref.~\cite{GN_RSFP}), 
as it can provide a very good fit to the solar neutrino data. 
Fixing the shape of the magnetic field profile, 
a fit to the solar neutrino data~\cite{solarnuexp} 
is performed by varying $\Delta m^2$ and 
the overall normalization of the magnetic field, or 
the average value, $\langle B \rangle$. 

\vglue -0.19cm 
%%%%%%%%%%%%%%%%%%%%%%%%%%%%%%%%%%%%%%%%%%%%%%%%%%%%%%%%%%%%
\hglue -0.8cm 
\psfig{file=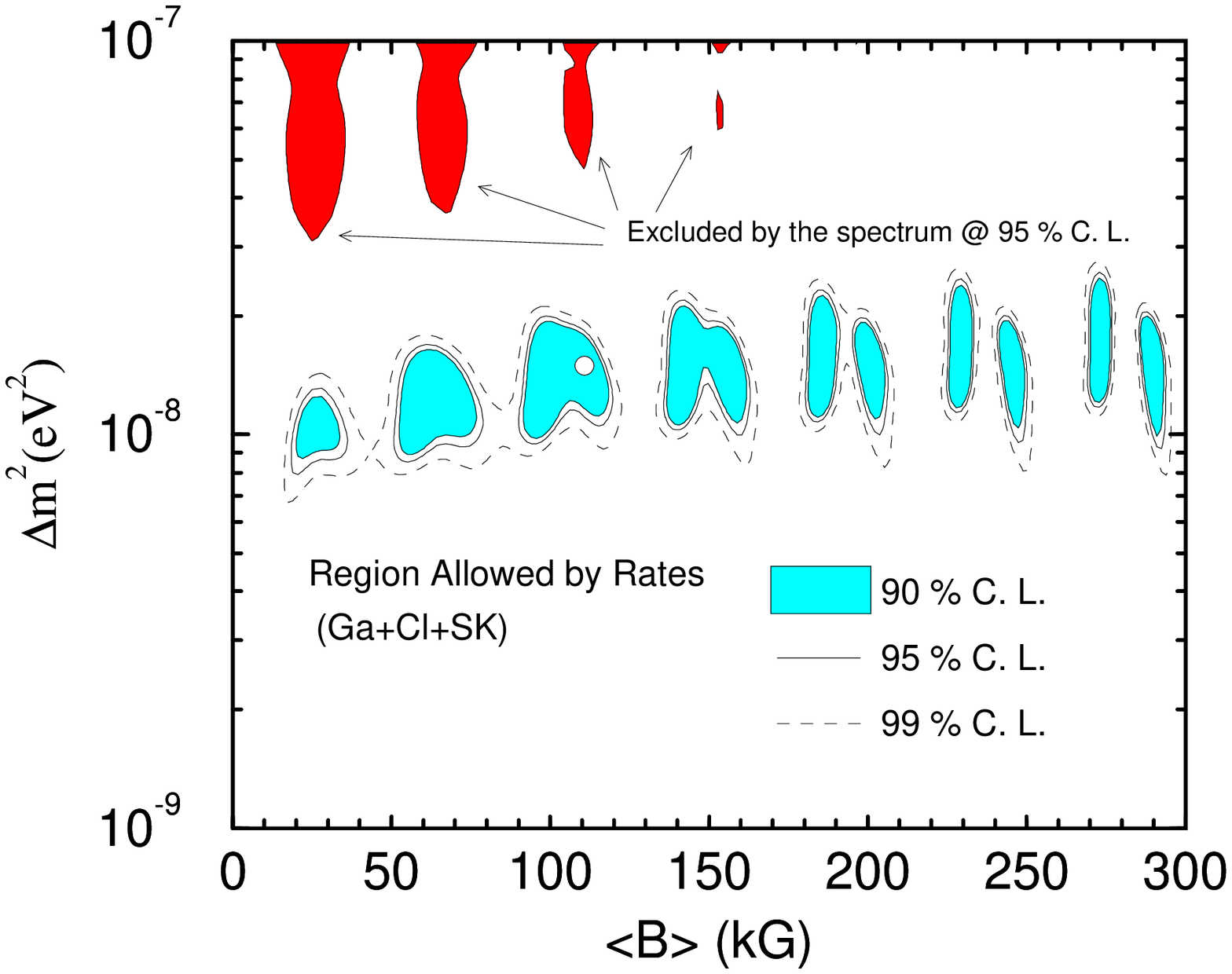,height=5.9cm,width=6.2cm}
\vglue -5.96cm 
\hglue 5.3cm 
\psfig{file=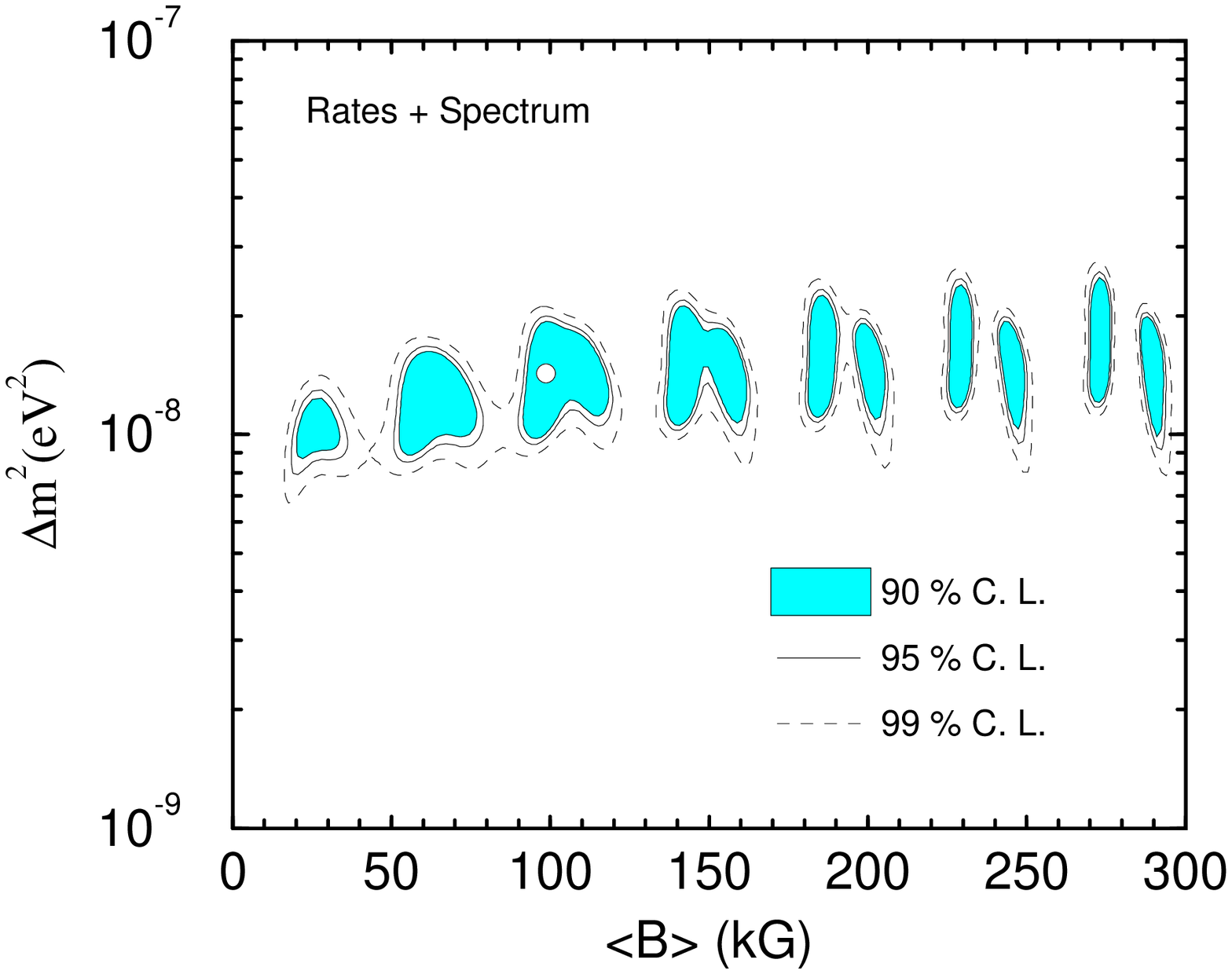,height=5.9cm,width=6.2cm}
\vglue -0.24cm 
\noindent
{\small Fig.\ 1: 
In the left panel the parameter region allowed by the rates as 
well as the excluded parameter region by the SK spectrum 
are presented and in the right panel the combined allowed
region is shown.}
%%%%%%%%%%%%%%%%%%%%%%%%%%%%%%%%%%%%%%%%%%%%%%%%%%%%%%%%%%%%
%
\newpage 
I present in Fig.\ 1 the region allowed by total rates
only, the Super-Kamiokande (SK) spectrum only and 
the combined case. 
For definiteness, I fix the magnitude of neutrino magnetic 
moment $\mu_\nu$ to $10^{-11} \mu_B$, where $\mu_B$ 
is the Bohr magneton, which is still allowed
by laboratory experiments~\cite{PDG00}.
Note, however, that the same results are obtained for 
different values of $\mu_\nu$ as long as the product 
$\mu_\nu \langle B \rangle$ is kept to be the same value. 
It is found that within the allowed region from the total 
rates, no strong distortion of the spectrum is expected, which 
is in good agreement with the current experimental situation. 
I show in Fig.\ 2 the expected SK recoil electron spectrum 
using the best fitted
%
%%%%%%%%%%%%%%%%%%%%%%%%%%%%%%%%%%%%%%%%%%%%%%%%%%%%%%%%%%%%
\vglue 0.1cm 
\hskip -1.0cm
\psfig{file=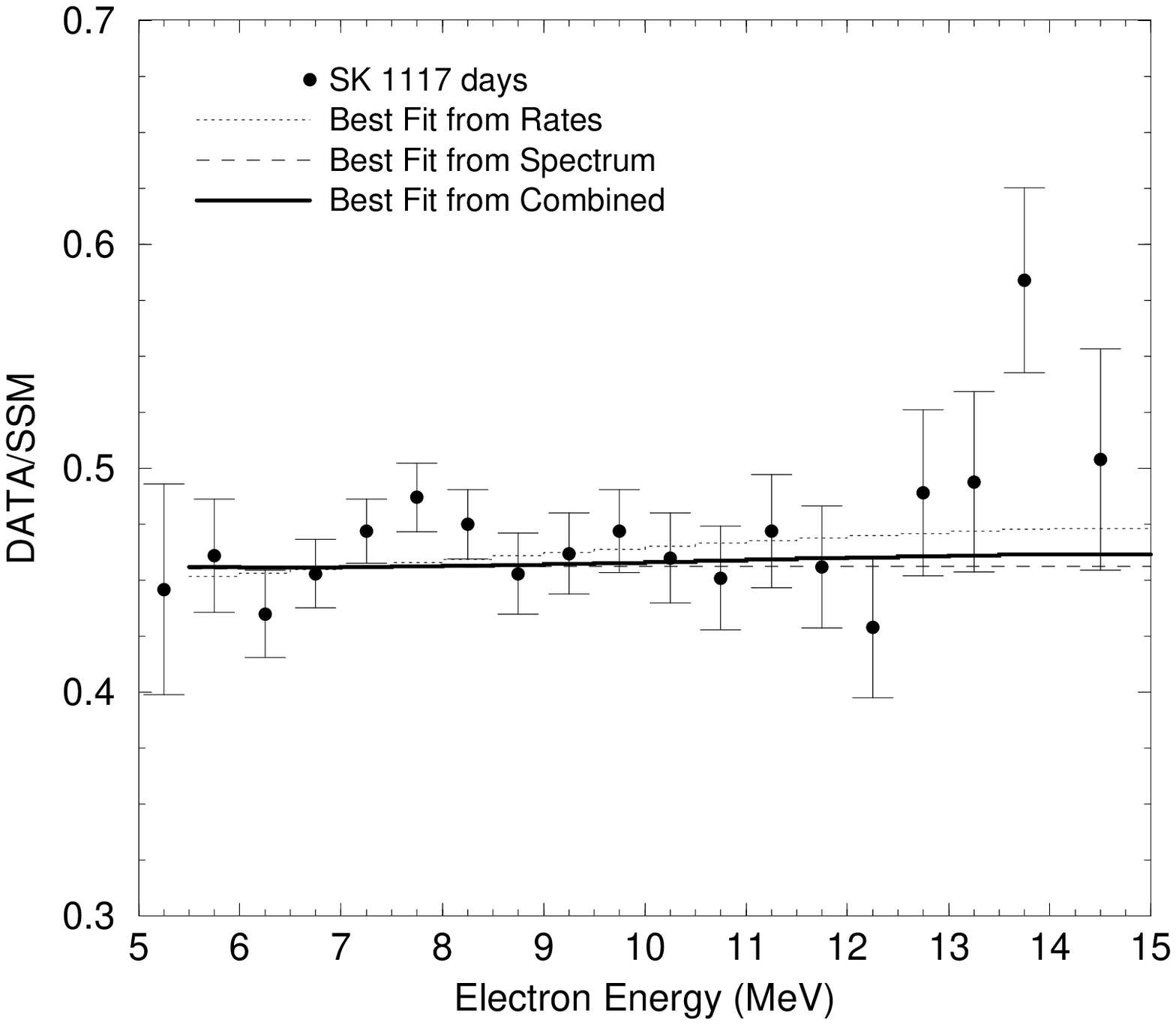,height=5.5cm,width=5.8cm}
{\hglue -5.6cm
\parbox[t]{5.2cm}
{\small Fig.\ 2: Expected SK recoil electron spectrum using 
the best fitted parameters from the total rates, spectrum 
only and combined analysis.}
}
%%%%%%%%%%%%%%%%%%%%%%%%%%%%%%%%%%%%%%%%%%%%%%%%%%%%%%%%%%%%
\vglue -6.8cm 
\hglue 5.0cm \parbox[t]{6.3cm}{
parameters. We see that no strong
distortion is expected. 
It is found that for the total rates, $\chi_{min}^2=0.17$ for 
2 degrees of freedom (DOF) and for the spectrum, 
$\chi_{min}^2=13.6$ for 15 DOF, and for the combined 
$\chi_{min}^2=18.3$ for 24 DOF, 
where the contribution from SK zenith angle dependence 
(6 bins) was included though the RSFP mechanism does not 
induce any significant distortion of the zenith angle
dependence. 
I conclude that RSFP mechanism can explain well the solar neutrino 
data provided that the product of neutrino magnetic moment
and the strength of the solar convective zone magnetic 
field satisfies $\mu_\nu \cdot \langle B \rangle 
\gsim 10^{-11} \mu_B \cdot O(10)$ kG. 
} 

\vspace{-0.1cm}
\section{Non-Standard Neutrino Interactions}

Next let me consider the solution based on resonant 
conversion induced by non-standard neutrino interactions
(NSNI) in matter~\cite{GMP,fcsol2,roulet,wol,valle87,bergmann}. 
Here I will consider the phenomenological approach by simply 
assuming the existence of a tree-level process 
$\nu_\alpha+ f \to \nu_\beta + f$ 
with an amplitude 
$\epsilon_{\alpha\beta}\sqrt{2} G_F$, 
where $\alpha$ and $\beta$ are flavor
indices, $f$ stands for the interacting elementary fermion 
($d$, $u$ quark or electron) and $\epsilon_{\alpha\beta}$ is 
considered to be free parameter, which characterize the strength of NSNI. 
Here I consider such NSNI induced only by $d$ or $u$-quark 
since if they are induced only by electrons, no resonant 
conversion can occur and the fit to the total rates
is not so good~\cite{fc_ele}. 

In the presence of such NSNI 
neutrino evolution equation in matter for the system of 
two massless neutrinos, $\nu_e -\nu_x (x=\mu,\tau$), 
is given as~\cite{GMP,fcsol2}:
\vspace{-0.1cm}
\begin{eqnarray} &  i{\displaystyle{d}\over 
\displaystyle{dr}}\left[ 
\begin{array}{c} \nu_e 
\\ \nu_x  \end{array} \right] = \hskip .3cm & \hskip-.1cm
\sqrt{2}\,G_F \left[ \begin{array}{cc} n_e(r) &  \epsilon n_f(r)
\\ \epsilon n_f(r)& \epsilon ' n_f(r) \end{array} \right]
\left[ \begin{array}{c} \nu_e  \\ \nu_x 
\end{array} \right] ,
\label{motion} 
\end{eqnarray}
\noindent
where, %I assumed neutrinos are massless (or have equal masses),  
$\epsilon \equiv \epsilon_{e x }$ and 
$\epsilon' \equiv \epsilon_{xx} - \epsilon_{ee}$.
Due to the presence of $\epsilon'$ term, a MSW-like resonant 
conversion can occur (when 
$n_e(r) = \epsilon'n_f(r)$ is satisfied) even if neutrinos
are massless~\cite{GMP,valle87}. 
The crucial point which makes this mechanism
a viable solution to the SNP~\cite{BGHKN}, despite 
the fact that the conversion probability itself 
is completely energy independent (see Eq. (\ref{motion})), 
is that different production distributions of neutrinos
can lead to different survival probability 
at the solar surface, after experiencing the resonance~\cite{GMP}.

Here I show some updated results~\cite{now2000_fc} 
of our previous fit~\cite{BGHKN}.
I present in Fig.\ 3 and 4 the allowed parameter region 
by the rates only and by the combined fit of 
rates, SK zenith angle dependence and SK spectrum assuming
NSNI with $d$-quark and $u$-quark, respectively. 
%
%%%%%%%%%%%%%%%%%%%%%%%%%%%%%%%%%%%%%%%%%%%%%%%%%%%%%%%%%%%%
%
\vglue - 0.15cm
\hglue -1.2cm 
\psfig{file=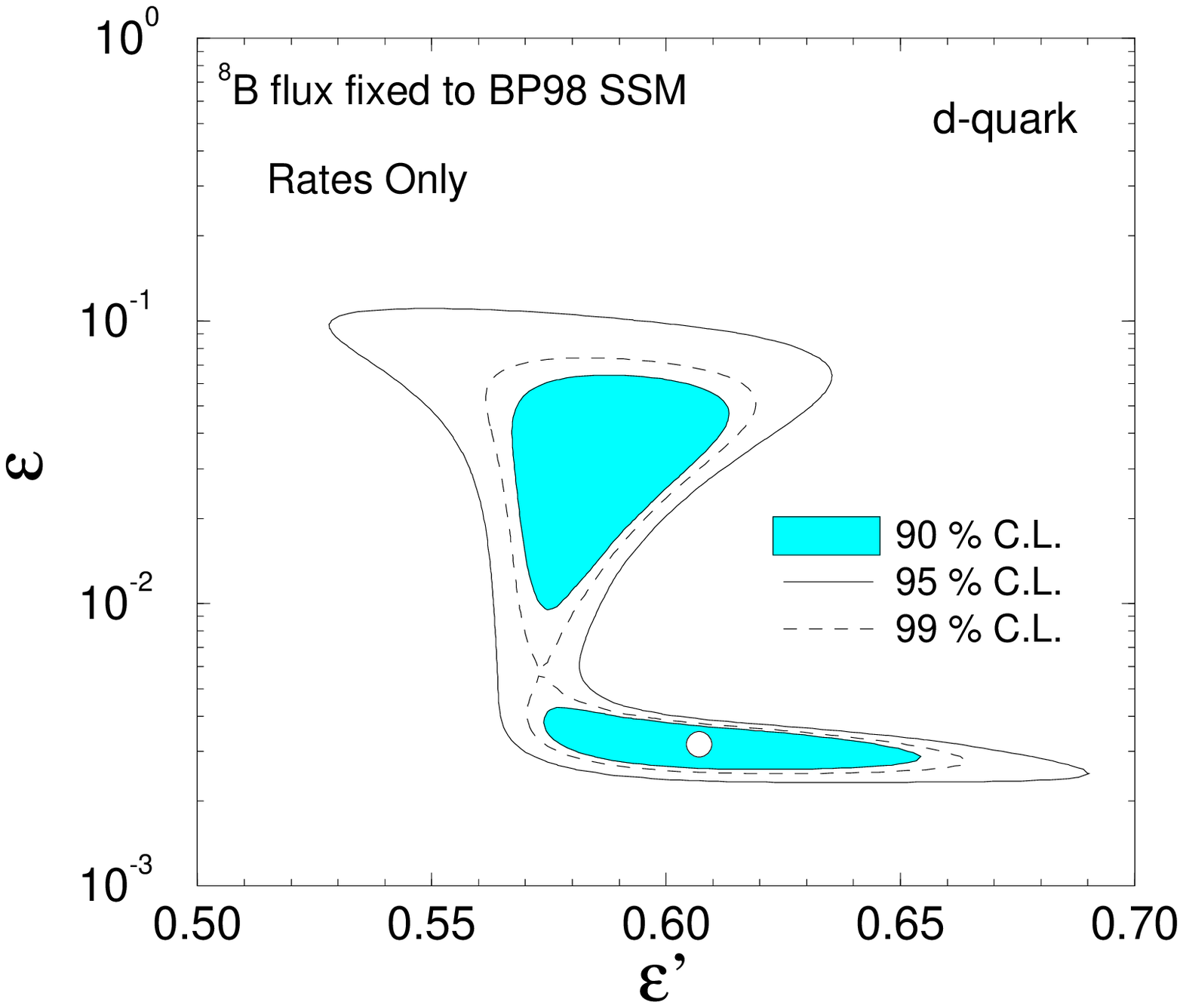,height=5.8cm,width=6.8cm}
\vglue -5.82cm 
\hglue 5.1cm 
\psfig{file=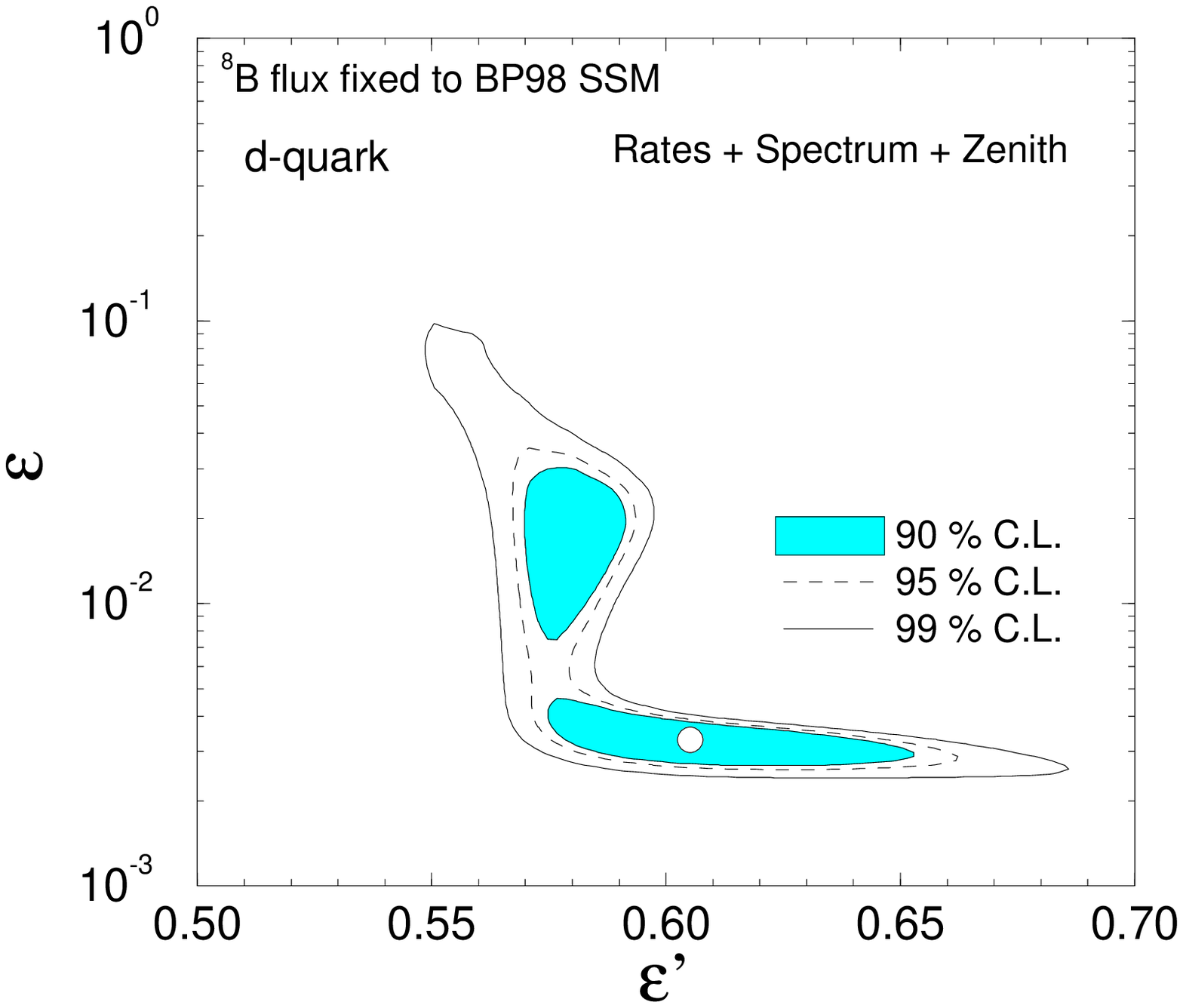,height=5.8cm,width=6.8cm}
\vglue -0.4cm 
\noindent
{\small 
Fig.\ 3: Allowed parameter region by the rates only 
(left panel) and the combined data (right panel)
assuming NSNI with d-quark. 
Best fit points are indicated by open circles. 
Adopted from Ref.~\cite{now2000_fc}. 
} 
\vglue -0.13cm 
%%%%%%%%%%%%%%%%%%%%%%%%%%%%%%%%%%%%%%%%%%%%%%%%%%%%%%%%%%%%
\hglue -1.2cm 
\psfig{file=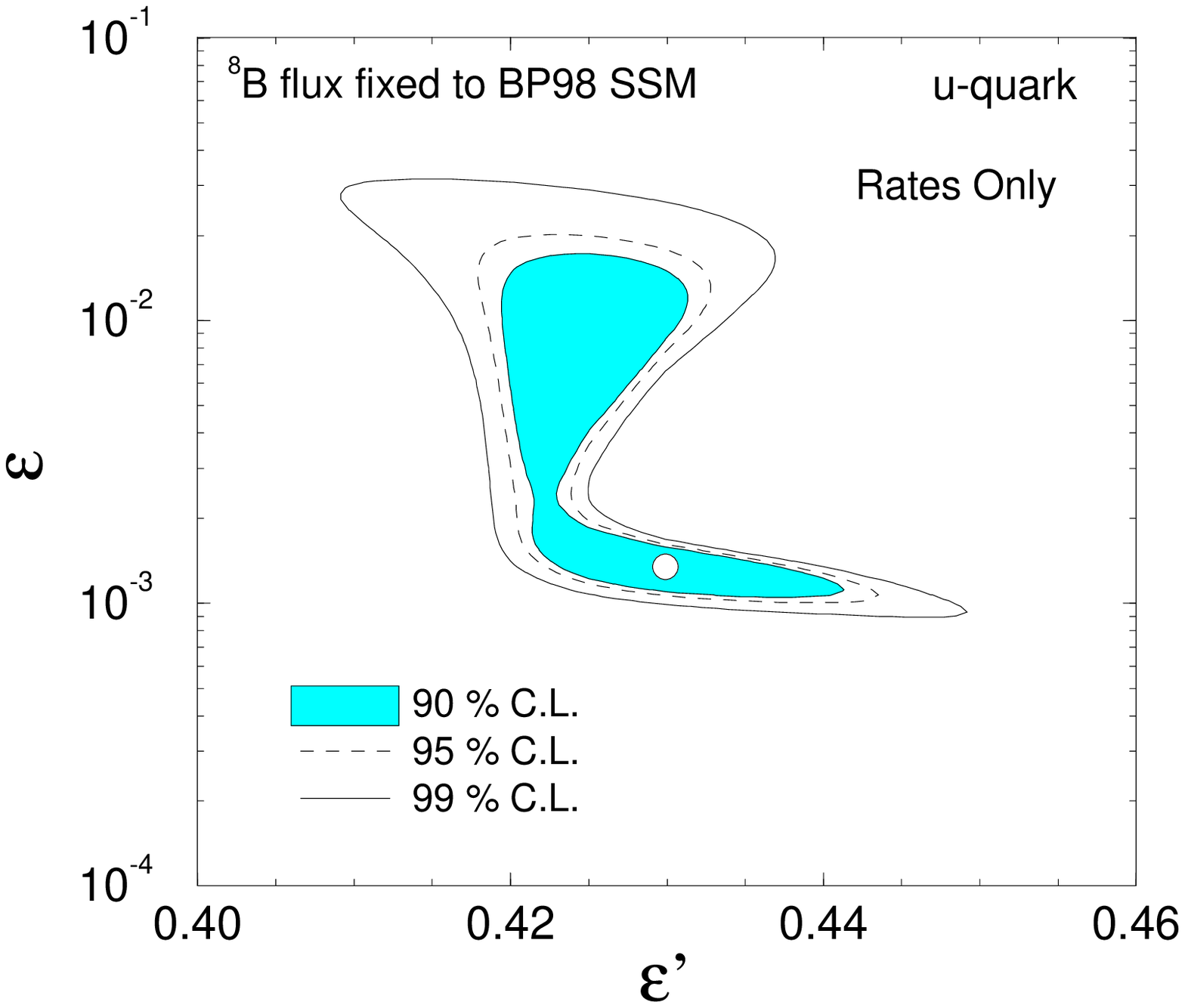,height=5.8cm,width=6.7cm}
%%%%%%%%%%%%%%%%%%%%%%%%%%%%%%%%%%%%%%%%%%%%%%%%%%%%%%%%%%%%
\vglue -5.88cm 
\hglue 5.1cm 
\psfig{file=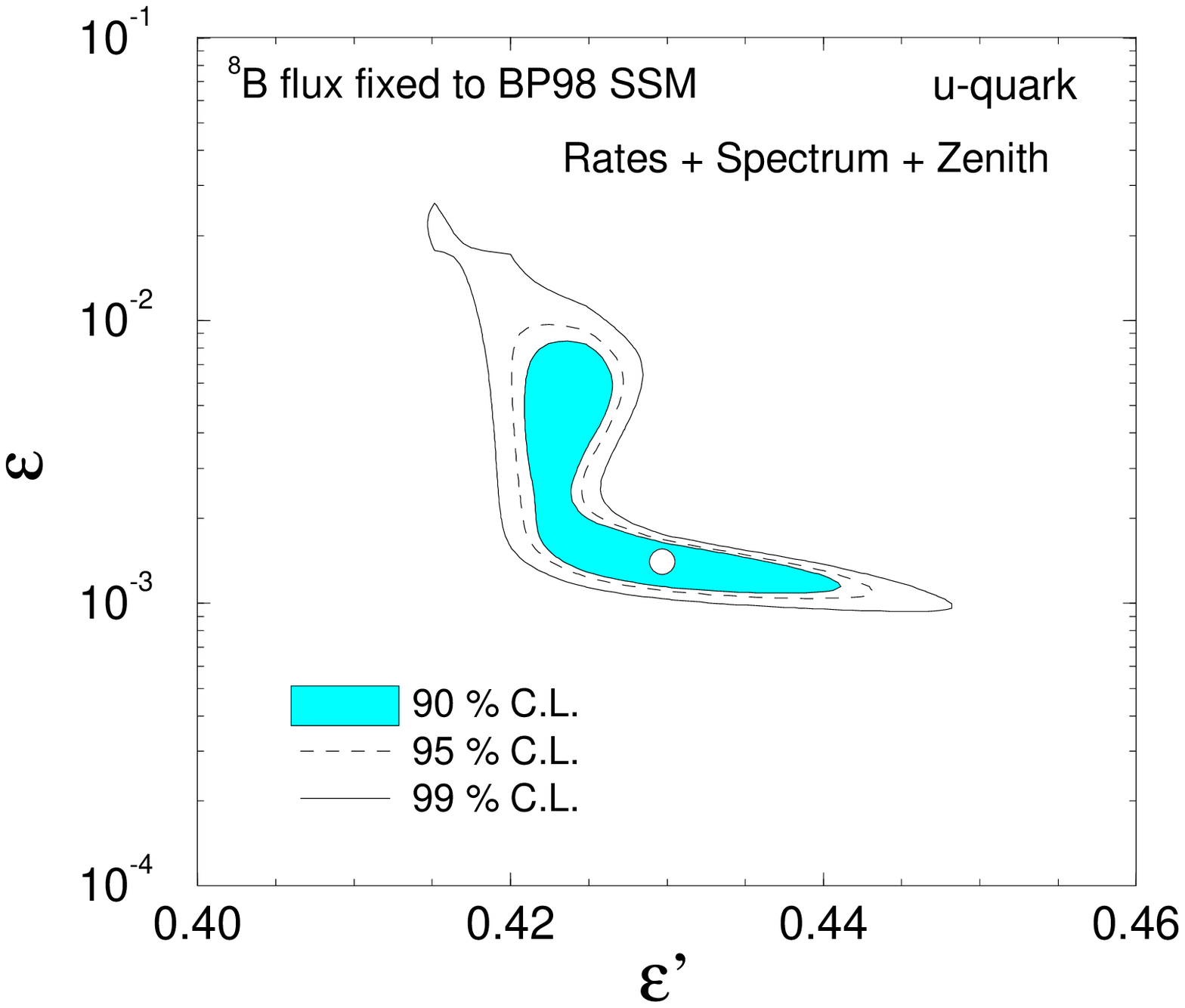,height=5.8cm,width=6.7cm}
\vglue -0.35cm 
\noindent
{\small 
Fig.\ 4: Same as in Fig.\ 3 but for u-quark interactions.
Adopted from Ref.~\cite{now2000_fc}. 
}
%%%%%%%%%%%%%%%%%%%%%%%%%%%%%%%%%%%%%%%%%%%%%%%%%%%%%%%%%%%%
%

It is found that $\chi^2 = 1.67$ for 2 DOF for the rates only 
and $\chi^2 = 19.6$ for 24 DOF for the combined fit for 
the case of $d$-quark and $\chi^2 = 1.68$ for 2 DOF for 
the rates only and $\chi^2 = 19.7$ for 24 DOF for the combined 
fit for the case of $u$-quark. 
I conclude that the solar neutrino
data can be well accounted for by this mechanism. 
In Ref.~\cite{BGHKN} it is discussed that values of 
$\epsilon$ and $\epsilon'$ required to have a good fit 
are still allowed by the laboratory experiments. 

\vspace{-0.2cm}
\section{Violation of Equivalence Principle}

Finally, let me consider a even more exotic scenario. 
The interesting idea that gravitational forces may induce 
neutrino mixing and flavor oscillations, if there 
exist violation of equivalence principle (VEP), 
was proposed  about a decade ago~\cite{gasperini}, 
and thereafter, 
many authors have investigated the possibility of solving 
%the SNP by such gravitationally, or VEP induced 
the SNP by such VEP induced neutrino 
oscillations~\cite{vep_msw,vep_us}. 
It is known that the solution to the SNP can be provided either
by the VEP induced MSW-like resonant conversion~\cite{vep_msw}
or by the VEP induced vacuum oscillation~\cite{vep_us}. 
Here, I consider only the latter solution
since the parameter required for the former one can 
be excluded by a laboratory experiment~\cite{pkm}. 
I will show some updated results~\cite{now2000_vep} 
of our previous analysis~\cite{vep_us}. 

Following the framework proposed in Refs.~\cite{gasperini}, 
to describe the VEP induced massless neutrino oscillation,  
phenomenologically, we can simply do 
the following replacement in the usual mass induced 
oscillation formula:
$\Delta m^2/2E \to 2E|\phi \Delta \gamma|$ 
and $\theta  \to \theta_G$, 
where 
$\Delta m^2$ is the mass squared difference, 
$\phi$ is the gravitational potential which is assumed 
to be constant in our work as it may come from 
the local Super-cluster~\cite{kenyon}, $\theta$ is the usual mixing angle 
which relate weak and mass eigenstates 
and $\theta_G$ is the mixing which 
relates weak and gravitational eigenstates, 
and  $\Delta \gamma$ is the quantity which measures 
the magnitude of VEP. 
See Ref.~\cite{veporigin} for some discussions on 
possible origins of VEP. 

The distinctive feature of this oscillation mechanism, compared
to the usual mass induced one, is that 
the oscillation wavelength $\lambda$ 
is inversely proportional to the neutrino 
energy, $\lambda \propto E^{-1}$. 
This energy dependence is very crucial in obtaining a good 
fit to the total rates without causing any problem 
to the fit of the SK spectrum~\cite{vep_us}, 
contrary to the situation in the case of usual mass 
induced vacuum oscillation solution to the SNP~\cite{VO3g}.  

Here I show some results~\cite{now2000_vep} updated 
from our previous analysis~\cite{vep_us}. 
I present in Fig.\ 5 the allowed parameter region 
in the $\sin^2 2\theta_G - |\phi \Delta \gamma|$ plane, 
determined only 
by the total rates (left panel) 
the allowed region determined by the SK spectrum only (right panel). 
In Fig.\ 6, in the left panel I present the allowed 
region for the combined fit of the rates and 
the spectrum and also I show in the right panel, 
the predicted spectra for the best fitted parameters, 
which are in good agreement with the data. 

It is found that $\chi^2_{min}$ = 1.78 for 2 DOF for 
the total rates and 
$\chi^2_{min}$ = 17.7 for 24 DOF for the combined 
analysis. 
I conclude that VEP induced long-wavelength oscillation 
can also provide a good fit to the solar neutrino data, 
provided that $\Delta \gamma \sim 10^{-20}$ assuming 
$\phi \sim 3 \times 10^{-5}$ for the local 
super-cluster~\cite{kenyon}.  
%
%%%%%%%%%%%%%%%%%%%%%%%%%%%%%%%%%%%%%%%%%%%%%%%%%%%%%%%%%%%%
\vglue -1.6cm 
\hglue -1.5cm 
\psfig{file=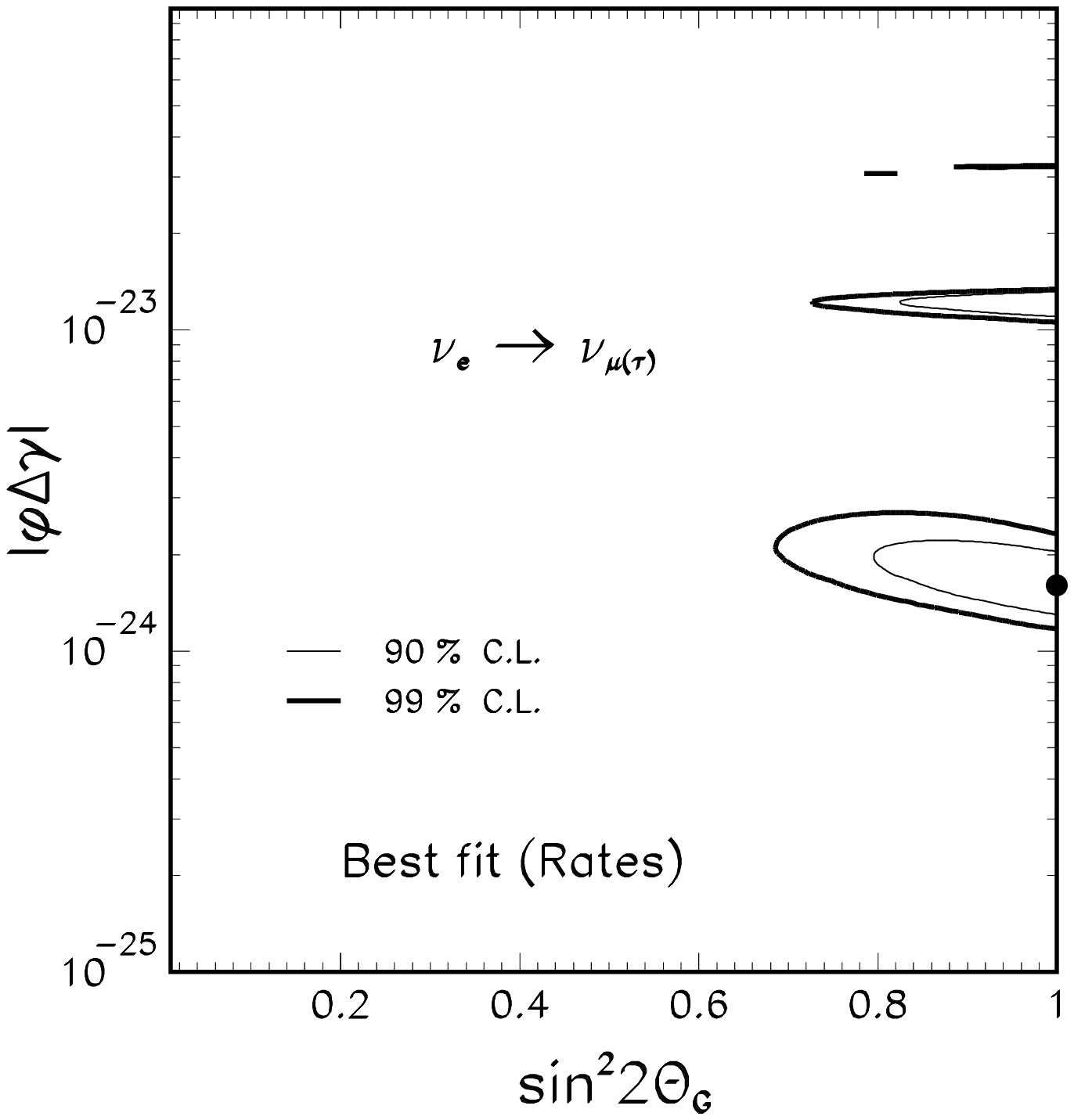,height=9.4cm,width=7.3cm}
%%%%%%%%%%%%%%%%%%%%%%%%%%%%%%%%%%%%%%%%%%%%%%%%%%%%%%%%%%%%
\vglue -9.69cm 
\hglue 4.7cm 
\psfig{file=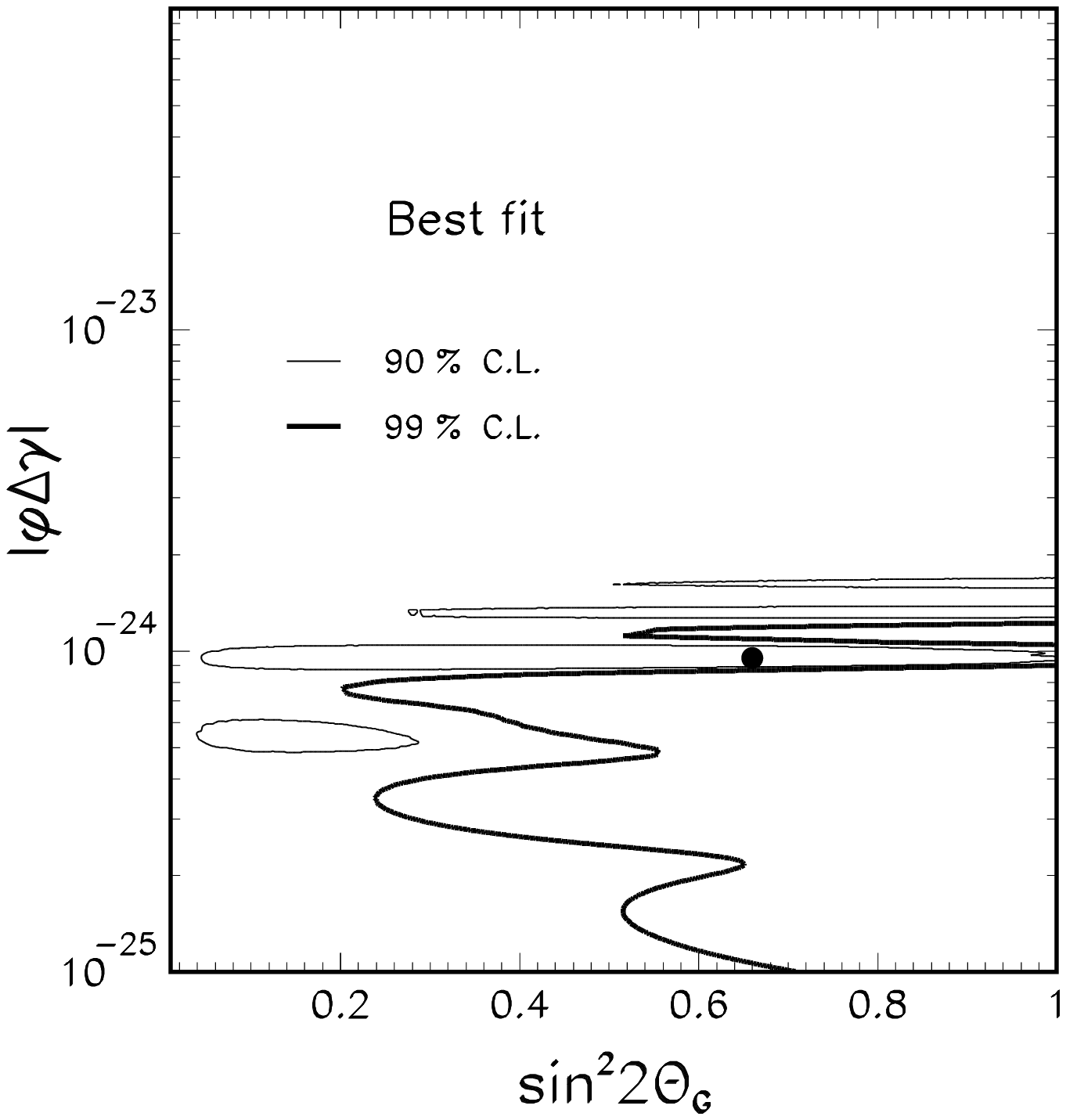,height=9.4cm,width=7.3cm}
\vglue -2.5cm 
\noindent
{\small 
Fig.\ 5: 
The left panel shows the allowed parameter region 
by the total rates only and the right panel shows
the allowed region by the SK recoil electron spectrum only 
where it is the inner part of the 
contours which is excluded by the SK spectrum. 
The best fit points are indicated by the filled circles.
Adopted from Ref.~\cite{now2000_vep}. 
} 
\label{fig1}
\vglue -1.6cm 
%%%%%%%%%%%%%%%%%%%%%%%%%%%%%%%%%%%%%%%%%%%%%%%%%%%%%%%%%%%%
\hglue -1.5cm 
\psfig{file=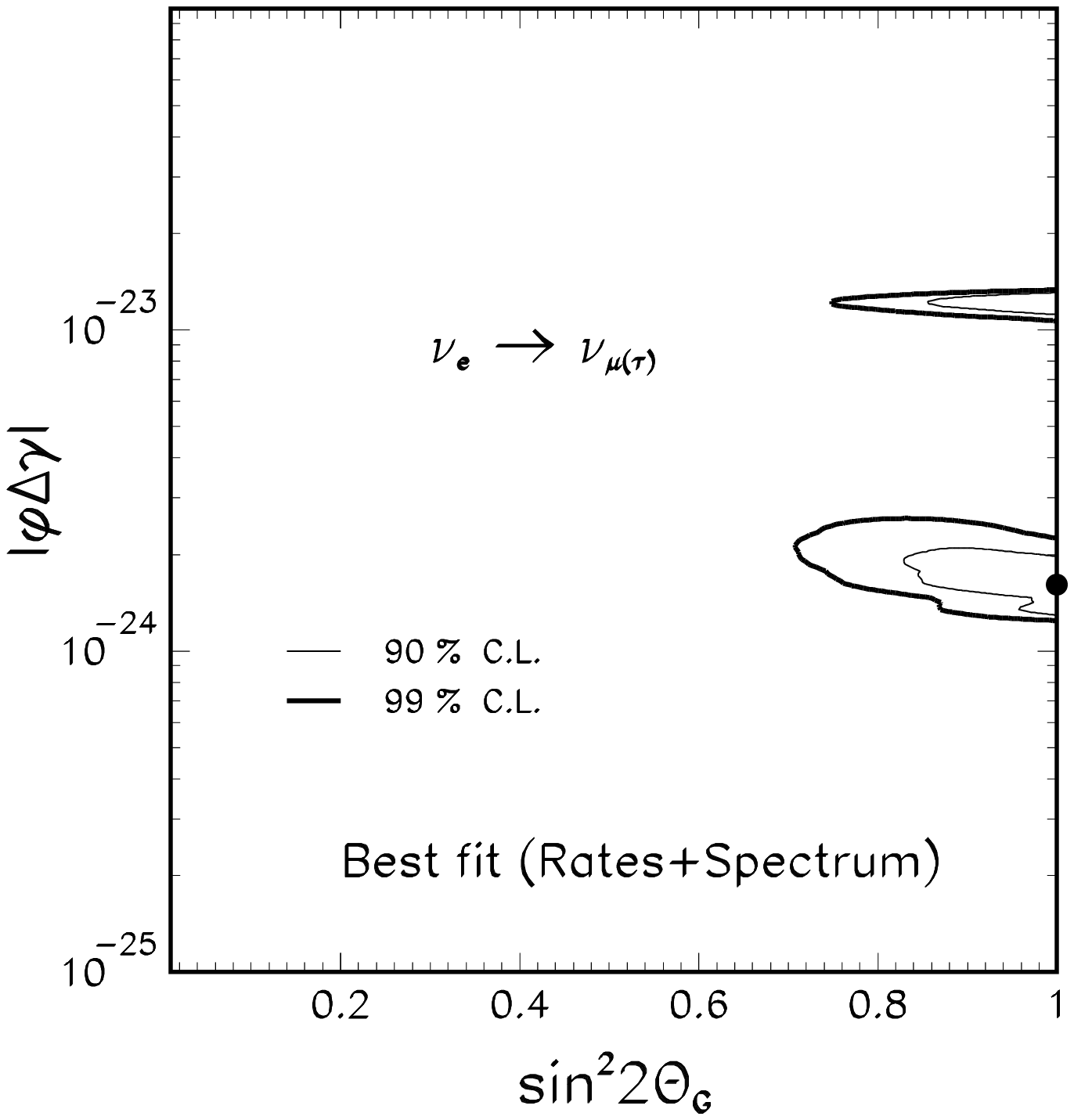,height=9.6cm,width=7.5cm}
\vglue -9.79cm 
\hglue 4.7cm 
\psfig{file=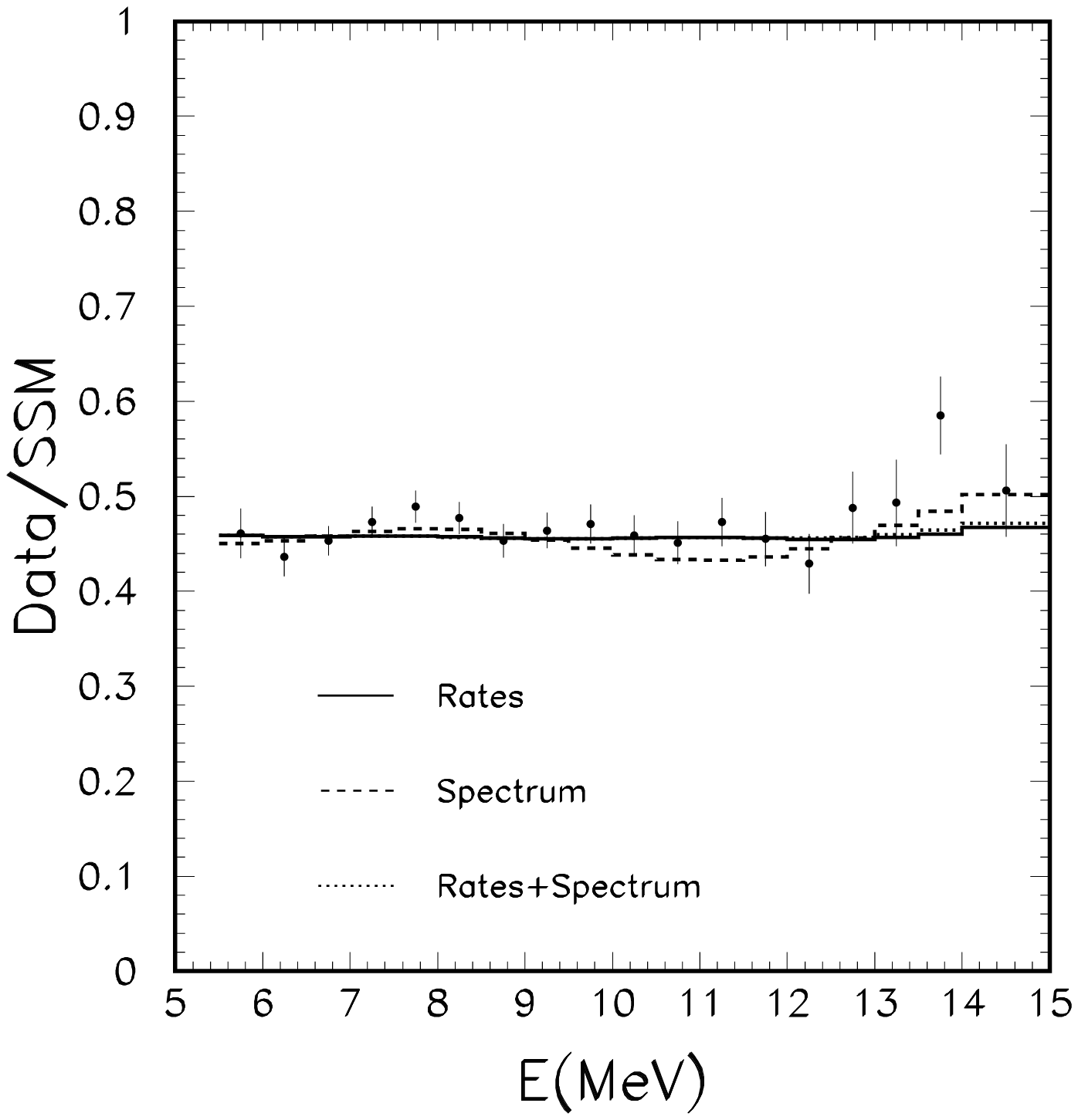,height=9.6cm,width=7.5cm}
\vglue -2.6cm 
\noindent
{\small 
Fig.\ 6: The left panel is the same as in Fig.\ 5 but for the 
rates and SK spectrum combined and the right panel shows
the expected SK spectrum for the best fit points. 
Adopted from Ref.~\cite{now2000_vep}. 
}
\vglue 0.2cm 
\label{fig3}
%%%%%%%%%%%%%%%%%%%%%%%%%%%%%%%%%%%%%%%%%%%%%%%%%%%%%%%%%%%%
%

\section{Implications for Low Energy Solar Neutrinos}

Let me now consider some simple implications for 
low energy solar neutrinos experiments, such as 
Borexino~\cite{borexino}, KamLAND~\cite{kamland}, 
Heron~\cite{heron}, Xmass (Xenon)~\cite{xenon} and 
GENIUS~\cite{genius} 
which aim to observe $^7$Be and/or $pp$ neutrinos
by means of $\nu_e e^-$ scattering reactions. 

Let me first define the following quantity, 
\vspace{-0.1cm}
\begin{equation}
R(X) \equiv 
\frac{\mbox{expected}\ \# \ \mbox{of}\ X\ \nu\ 
\mbox{events w oscillation}}
{\mbox{expected} \# \ \mbox{of}\ X\ \nu\ \mbox{events w/o oscillation}}
\ \ , 
\vspace{-0.1cm}
\end{equation}
\noindent
where $X = ^7$Be or $pp$. 
Explicitly, $R(X)$ can be given by, 
\vspace{-0.1cm}
\begin{equation}
R(X) \equiv
\frac{ \displaystyle \int dE_e 
\int dE_\nu \phi_{X}(E_\nu) 
\left( 
\frac{d \sigma_{\nu_e}}{dE_e}
P_{ee}(E_\nu) 
+ 
\frac{d \sigma_{\nu_\alpha}}{dE_e} 
\left[ \ 1-P_{ee}(E_\nu) \ \right]
\right) 
}
{\displaystyle 
\int dE_e \int dE_\nu 
\phi_{X}(E_\nu) 
\frac{d \sigma_{\nu_e}}{dE_e} 
},
\label{eq:rates2}
\vspace{-0.1cm}
\end{equation}
\noindent
where $E_e$ is the recoil electron energy, 
$\phi_{X}(E_\nu)$ is the neutrino energy 
distribution of $X$ solar neutrinos,  
$ d \sigma_{\nu_{e,\alpha}}/{dE_{e}}$ is 
$\nu_{e,\alpha} e^- $ $(\alpha=\mu,\tau)$ 
scattering cross sections
and $P_{ee}(E_\nu)$ is the $\nu_e$ survival probability. 
For simplicity, I do not take into account 
the resolution function in this contribution.

Next let me try to derive some relation that $R(pp)$ and 
$R(^7\mbox{Be})$ must satisfy. 
The only ongoing experiment which can detect these neutrinos 
(but not separately) is the $^{71}$Ga experiment. 
The contribution of neutrinos to $^{71}$Ga experiment 
from different reaction sources is expressed as~\cite{BP98},
\vspace{-0.1cm}
\begin{equation}
S_{Ga}  \simeq 69.6 \langle P_{ee}(pp) \rangle 
        + 34.4 \langle P_{ee}(^7\mbox{Be}) \rangle 
	+ 12.4 \langle P_{ee}(^8\mbox{B}) \rangle + ....
\ \ 	\mbox{SNU},
%        +...(minor contributions)
\vspace{-0.1cm}
\end{equation}
\noindent
where $\langle P_{ee}(X) \rangle\ (X=pp,^7\mbox{Be},^8\mbox{B})$ 
indicate the survival probabilities of $X$ neutrinos and 
I ignored some other minor contributions. 
Using the relations, 
$R(X) \sim \langle P_{ee}(X) \rangle
+ r(X) [1-\langle P_{ee}(X) \rangle]$, 
where $r(X)$ denotes the ratio of the cross sections 
$\langle \sigma_{\nu_{\mu,\tau} e}\rangle / 
\langle \sigma_{\nu_e e}\rangle$ 
appropriately averaged over 
the energy spectrum of $X$ neutrinos, 
and the observed results of   $^{71}$Ga experiment, 
$S^{obs}_{Ga}\simeq 75 (1\pm 0.1) $ SNU,   
I obtain, 
\vspace{-0.1cm}
\begin{equation}
R(^7\mbox{Be})+ 2R(pp) \sim 2 \pm 0.2.
\label{relation_low}
\end{equation}
\vspace{-0.1cm}
\noindent
This is the condition which must be satisfied by any acceptable 
solution in order to account well for the measurement of 
the $^{71}$Ga experiment.

Now let me try to predict the ranges of $R(pp)$ and 
$R(^7\mbox{Be})$ for the various solutions to the solar neutrino 
problem I discussed in the previous sections. 
I try to ``map'' the 95 \% C.L. allowed parameter region 
of each solutions into the plane spanned by $R(pp)$ 
and $R(^7\mbox{Be})$.  
 I present in Fig.\ 7 the expected range of 
$R(pp)$ and $R(^7\mbox{Be})$ for various solutions.  
For the purpose of comparison, I also plot the expected range
for the standard mass induced oscillation solutions, namely, 
MSW large mixing angle (LMA), MSW small mixing angle (SMA), 
MSW low-$\Delta m^2$ (LOW)~\cite{GHPV,review}
and vacuum oscillation (VAC) solutions~\cite{VO3g}
to the SNP.
We can confirm that, roughly speaking, all the solutions
satisfy the above relation in Eq. (\ref{relation_low}). 

%%%%%%%%%%%%%%%%%%%%%%%%%%%%%%%%%%%%%%%%%%%%%%%%%%%%%%%%%%%%
\vglue -0.4cm
\hglue -3.0cm
\centerline{
\psfig{file=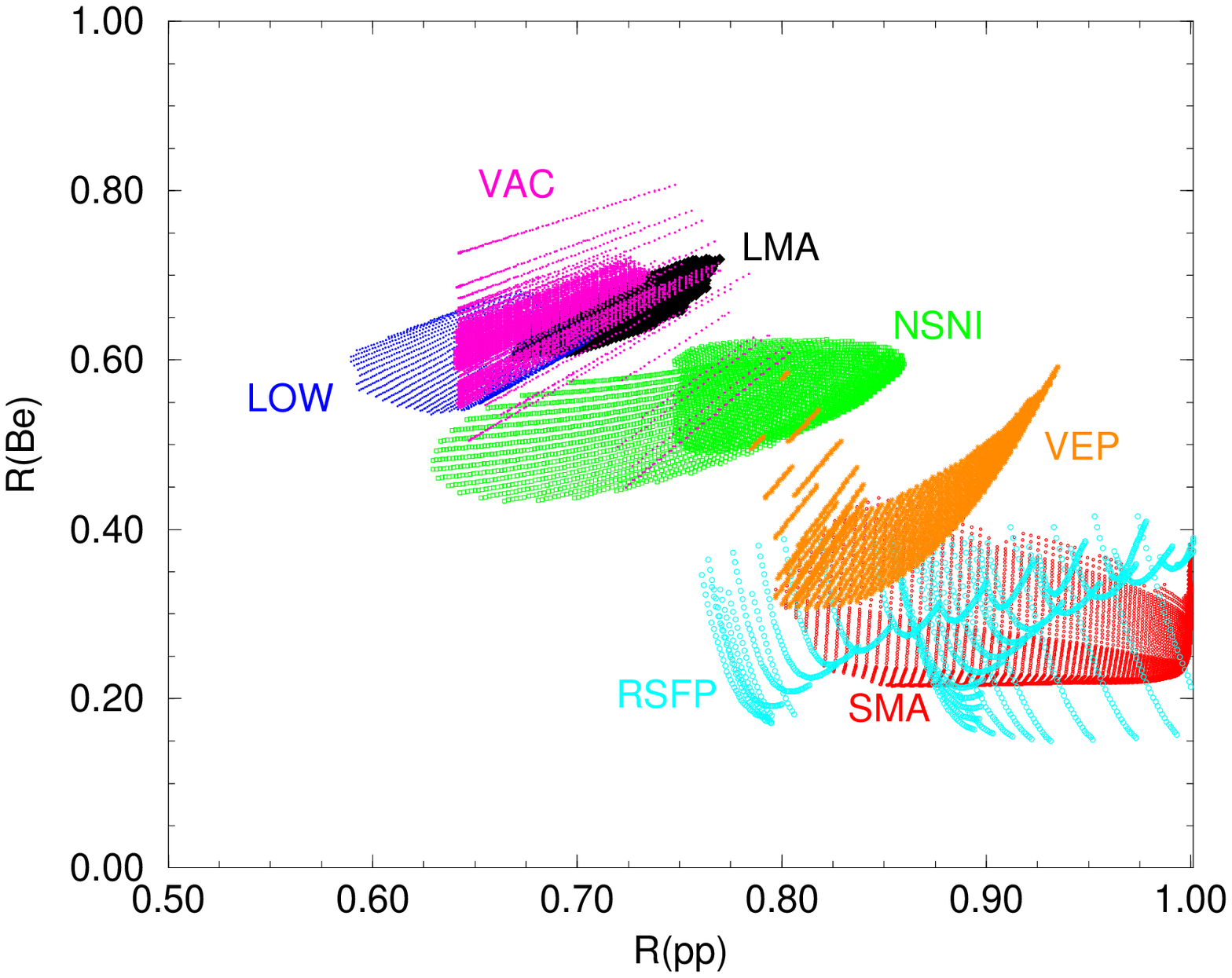,height=8.1cm,width=8.2cm}
}
%\vglue -4.5cm 
% \noindent
%\hglue 8.0cm 
\vglue -0.5cm 
\hglue -0.5cm 
\parbox[]{6.9cm}
{\small 
Fig.\ 7: Predictions for the total rates for 
$pp$ and $^7$Be neutrinos for various solutions, 
which are determined from the 95 \% C.L. allowed
parameter regions. }
%%%%%%%%%%%%%%%%%%%%%%%%%%%%%%%%%%%%%%%%%%%%%%%%%%%%%%%%%%%%
\vglue -8.6cm 
\hglue 6.9cm 
\parbox[]{4.5cm}
{
\ \ \ 
As far as rates of $pp$ and $^7$Be neutrinos are concerned, 
LMA, LOW and VAC solutions have significant overlap. 
Similarly, RSFP and SMA have also large overlap. 
On the other hand, VEP and NSNI solutions have less
overlap with the others. 
From these observations, I can say that 
LMA, LOW and VAC solutions could be easily 
confused and the same applies to RSFP and SMA solutions
if we will use only the information of $R(pp)$ 
and $R(^7\mbox{Be})$.
By combining some other information such as zenith angle dependence, 
time variations,
or spectrum distortion, etc.,
we can discriminate some of 
}
\vglue 0.11cm
{\noindent  
these solutions from the standard mass induced oscillation 
solutions~\cite{GN_RSFP,BGHKN,vep_us,review}. 
}

\vspace{-0.3cm}
\section{Conclusions}

I showed that alternative solutions to the solar neutrino problem, 
which do not invoke neutrino mass and/or flavor mixing,  
still exist and they can provide as good fit as the 
standard oscillation explanations such as the MSW and VAC
solutions to the SNP. 
While these non-standard solutions are theoretically less 
motivated than the standard ones which are based only 
on neutrino mass and mixing, 
let me stress that it would be important 
to exclude such non-standard solutions experimentally, 
in order to clearly establish the standard solutions 
such as the MSW ones. 
I also considered some possible implications for the 
future low energy solar neutrino experiments and sketched 
some general features focusing only 
on the total rates.
More detailed and careful comparisons of these solutions 
as well as some further implications for future solar neutrino 
experiments will be done elsewhere~\cite{review}. 

\vspace{-0.3cm}
\section*{Acknowledgments}
\vspace{-0.1cm}
I would like to thank the organizers of the workshop 
for invitation. 
I am grateful to S. Bergmann, 
A. M. Gago,  M. M. Guzzo, P. I. Krastev, P. C. de Holanda, 
R. Zukanovich Funchal for collaborations and 
D. V. Ahluwalia, V. Berezinsky, H. Minakata, 
C.\ Pe\~na-Garay, O. L. G. Peres, Y. Suzuki, J. W. F. Valle, 
F. Vissani for valuable discussions. 
I thank A. M. Gago, P. C. de Holanda and 
R. Zukanovich Funchal for their kind help 
in preparing this contribution. 
The author was supported by the Brazilian funding 
agency FAPESP.

\end{document}